\documentclass[twocolumn]{article}
\usepackage[letterpaper, hmargin=1.5cm, vmargin=2cm]{geometry}
\usepackage{algorithm}
\usepackage{algpseudocode}
\usepackage{url}
\usepackage{authblk}
\usepackage{amsmath,amsfonts,amssymb}
\usepackage{graphicx,xcolor}
\usepackage[section]{placeins}
\usepackage[switch]{lineno}
\newcommand{\hl}[1]{#1}
\providecommand{\keywords}[1]
{
	\small
  \textbf{\textit{Keywords---}} #1
}

\title{Fast generation of arbitrary optical focus array}

\author[1]{Xin Liu}
\author[1]{Yiwen Hu}
\author[1]{Shijie Tu}
\author[1]{Cuifang Kuang}
\author[1]{Xu Liu}
\author[1,*]{Xiang Hao}

\affil[1]{State Key Laboratory of Modern Optical Instrumentation, College of Optical Science and Engineering, Zhejiang University, Hangzhou 310027, China}
\affil[*]{Corresponding author: haox@zju.edu.cn}

\begin{document}

\maketitle

\begin{abstract}
	We report a novel method to generate arbitrary optical focus arrays (OFAs). Our approach rapidly produces computer-generated holograms (CGHs) to precisely control the positions and the intensities of the foci. This is achieved by replacing the fast Fourier transform (FFT) operation in the conventional iterative Fourier-transform algorithm (IFTA) with a linear algebra one, identifying/removing zero elements from the matrices, and employing a generalized weighting strategy. On the premise of accelerating the calculation speed by $>$70 times, we demonstrate OFA with 99\% intensity precision in the experiment. Our method proves effective and is applicable to systems in which real-time OFA generation is essential.
\end{abstract}


\section{Introduction}
Optical focus arrays (OFAs) are useful in optical trapping~\cite{Barredo:2018:Nature}, laser fabrications~\cite{Li:2020:LaserPhotonicsRev.,He:2022:LaserPhotonicsRev.}, optogenetics~\cite{Packer:2015:Nat.Methods,Yang:2021:Science}, high-throughput microscopy~\cite{Xue:2019:Optica}, and holographic displays~\cite{Ren:2020:Sci.Adv.}. To generate an OFA, one may opt for the use of microlens arrays~\cite{Tang:2011:Appl.Opt.,Pang:2012:Opt.Lett.}, acoustic-optic deflectors~\cite{Endres:2016:Science}, Dammann gratings~\cite{Garcia:2012:Appl.Opt.}, or spatial light modulators (SLMs)~\cite{Matsumoto:2012:Opt.Lett.,Lou:2013:Sci.Rep.,Li:2020:LaserPhotonicsRev.}. Among these, SLMs use programmable computer-generated holograms (CGHs) that allow flexibly generating OFAs and dynamically compensating for optical system imperfections. In particular, phase-only SLMs use optical power more efficiently than those amplitude modalities, which necessarily attenuate the incident light.

Conventional algorithms for creating such CGHs for the OFA generation using phase-only SLMs include non-iterative and iterative approaches. The non-iterative algorithms can generate the OFA at high speed, but they always encounter a severe trade-off between the spot number and the diffraction efficiency. In contrast, the iterative methods, i.e., iterative Fourier-transform algorithms (IFTAs), such as Gerchberg-Saxton (GS)~\cite{Gerchberg:1972:Optik}, weighted GS (GSW)~\cite{DiLeonardo:2007:Opt.Express}, and their variations~\cite{Kim:2019:Opt.Lett.,Li:2021:Front.Phys.}, are more preferred due to the capability of the large-scale OFA generation with high diffraction efficiency.

Fast Fourier transform (FFT) algorithm naturally compatible with IFTA~\cite{Cooley:1965:FFT}. However, the FFT-based implementation intrinsically imposes two undesirable constraints. One is the fixed sampling relationship between the pupil plane and the focal plane. This constraint always requires a zero-padding on the pupil plane and a cropping operation on the focal plane, respectively. This is aimed at reducing the large sampling interval of the focal plane and extracting the region of interest (ROI) from its broad spatial scope, which arises from the limited numerical aperture (NA) of the lens and the small sampling interval on the pupil plane. The second one is the unalterable uniform sampling interval on the two planes. It prevents the IFTA from the generation of flexibly arranged OFA, because, to generate an OFA with precise positions of each focal spot, the sampling interval of the focal plane should be as small as possible. It should be at least close to the greatest common divisor among spatial distances between every two adjacent focal spots. Unfortunately, restricted by the first constraint, this, in turn, requires enormous zero-padding on the pupil plane, intensely increasing the computational burden. The first constraint can be alleviated by generalizing the FFT to the chirp-Z transform (CZT)~\cite{Rabiner:1969:CZT}, which requires two FFTs and one inverse FFT of three small matrices per operation, whereas the second constraint remains. A more flexible method is to write the two-dimensional (2D) discrete Fourier transform (DFT) as the matrix triple product (MTP)~\cite{Jurling:2018:J.Opt.Soc.Am.A, Zhao:2020:Opt.Lett.}, in which both the two constraints of the FFT are released. Nevertheless, to the best of our knowledge, neither of them was demonstrated for the OFA generation.

To accelerate the convergence and enhance the uniformity of the OFA, a weighting process is often integrated into the IFTA~\cite{DiLeonardo:2007:Opt.Express}. For the OFA generation with various intensities of each focal spot, although it can be realized via the conventional GS-based IFTA without the weighting process, the precise control of the intensity remains challenging.

In this paper, we develop a method to rapidly generate arbitrary OFA. The conventional FFT-based diffraction modeling is replaced by the MTP, which tremendously improves the computational efficiency, as well as the flexibility. Moreover, we further develop a generalized weighting strategy to accelerate the convergence.

\section{Results and discussion}
In the IFTAs, the forward propagation of the light is modeled following the far-field diffraction, which can be described as

\begin{equation}
	\varTheta \left(x, y \right) = \iint \limits_{k_{x}^{2} + k_{y}^{2} \leq k_{c}^{2}} E \left(k_{x}, k_{y} \right) e^{-i2\pi \left(k_{x}x + k_{y}y\right) } dk_{x}dk_{y},
	\label{eq:FFT-forward-propagation}
\end{equation}
where $\varTheta$ is the electric field distribution on the focal plane, and $\left(x, y \right)$ are the corresponding lateral coordinates. $\left(k_{x}, k_{y}\right)$ are the coordinates on the pupil plane, represented in the spatial frequency domain, and $k_{c} =\frac{\mathrm{NA}}{\lambda}$ is the cutoff frequency of the lens ($\lambda$ is the light wavelength). $E = Ae^{i\phi}$ is the electric field distribution of the incident beam on the pupil plane, i.e., pupil function. $A$ is the amplitude, and $\phi$ is the CGH used for the OFA generation.

Conventionally, the discrete computation of Eq.~\ref{eq:FFT-forward-propagation} is implemented with the 2D FFT algorithm. However, it also can be expressed as the MTP form~\cite{Jurling:2018:J.Opt.Soc.Am.A, Zhao:2020:Opt.Lett.},

\begin{equation}
	\begin{aligned}
		\varTheta & = \varOmega_{y} E \varOmega_{x}               \\
		          & = \mathcal{P}_{\mathrm{MTP}} \left(E \right),
	\end{aligned}
	\label{eq:MTP-forward-propagation}
\end{equation}
where $\varOmega_{x} = e^{-i2\pi K_{x}^{T}X} $, and $\varOmega_{y} = e^{-i2\pi Y^{T}K_{y}}$. $K_{x}$, $K_{y}$, $X$, and $Y$ are the coordinates, represented by row vectors, in the spatial frequency domain and the spatial domain, respectively. $T$ represents the transpose. $\mathcal{P}_{\mathrm{MTP}}$ is the MTP operator for forward propagation, and its inverse form can be formulated similarly.

\begin{figure}[!b]
	\centering
	\vspace{-3pt}
	\includegraphics{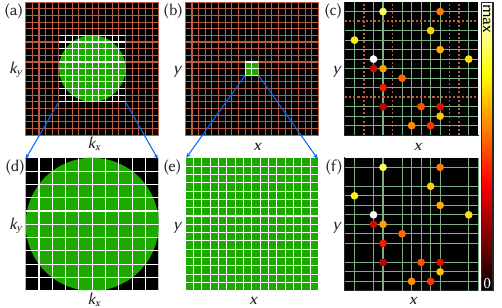}
	\caption{\label{fig:principles-of-MTP} Comparison between 2D FFT and MTP. (a) and (b) show the respective sampling grids on the pupil plane and focal plane of the 2D FFT. The orange grid represents the padded zeros of the pupil function or the calculated region with a near-zero intensity on the focal plane. The white grids represent the effective sampling ROI, in which the green areas illustrate the effective region of the incident field and focused field. (c) Uniform sampling of target OFA in conventional IFTA. The dashed orange lines indicate the redundant samplings. (d) and (e) are the two sampling grids for MTP, zoomed from (a) and (b), respectively. (f) Nonuniform sampling of target OFA in MTP.}
	\vspace{-3pt}
\end{figure}

Eq.~\ref{eq:MTP-forward-propagation} indicates that, in the MTP operation, the sampling intervals on both the pupil plane and focal plane can be nonuniform, although a uniform sampling interval is often required on the pupil plane as the practice for most of the modulation devices. Moreover, the fixed sampling relationship between the two planes is removed. Notably, the sampling requirements for avoiding aliasing still hold since this is the precondition of all the discrete implementations.

Figure~\ref{fig:principles-of-MTP} shows the superiority of MTP for arbitrary OFA generation. In the conventional IFTAs, the pupil function [Fig.~\ref{fig:principles-of-MTP}(a)] always requires the zero-padding to realize a fine sampling on the focal plane, but the ROI of which is only a small portion of the full spatial scope [Fig.~\ref{fig:principles-of-MTP}(b)]. Worse still, the focal plane is restricted by a uniform sampling interval, which inevitably results in many redundant samplings [Fig.~\ref{fig:principles-of-MTP}(c)]. In contrast, for the MTP, the zero-padding of the pupil function no longer exists [Fig.~\ref{fig:principles-of-MTP}(d)], and the ROI of the focal plane can be arbitrarily selected and sampled [Fig.~\ref{fig:principles-of-MTP}(e)], thereby eliminating the redundant samplings [Fig.~\ref{fig:principles-of-MTP}(f)].

The weighting approach in the GSW algorithm is very effective for the OFA generation with uniform intensity. However, it can not deal with the nonuniform case. To address this shortcoming, we generalize this weighting approach to make it compatible with the OFA generation with nonuniform intensity distribution. The details of our generalized weighting strategy cooperating with the MTP for arbitrary OFA generation are presented in Algorithm~\ref{algo:focus-array-generation}. In brief, we firstly give an initial guess of CGH and then calculate the complex amplitudes and intensities of foci using the forward MTP operator. The weighting factors are updated with the target amplitudes and the calculated ones of each focal spot, following which the spatial-domain amplitude constraint is applied. The pupil function is calculated via the inverse MTP operator and updated by applying the spatial-frequency-domain amplitude constraint. The iteration continues until the intensity accuracy reaches the threshold value.

\renewcommand{\algorithmicrequire}{\textbf{Input:}}
\renewcommand{\algorithmicensure}{\textbf{Output:}}
\newcommand{\Break}{\State \textbf{break} }
\newcommand{\idxn}{\left(n\right)}
\begin{algorithm}[!b]
	\caption{Fast generation of arbitrary OFA}\label{algo:focus-array-generation}
	\begin{algorithmic}
		\Require Amplitude of pupil function $A$. Initial guess of CGH $\phi_{0}$. Target intensity distribution of OFA $\widetilde{I}^{\idxn}, n = 1, 2, \dots, N$ ($N$ is the total number of focal spots). Initial weighting factors of each focal spot $w_{0}^{\idxn} = \left\{1, 1, \dots, 1\right\}$. Initial step count $j = 0$. Threshold value of intensity accuracy for termination $\mathcal{X} $.
		\Ensure CGH $\phi $ for desired OFA generation.

		\State Compute initial pupil function $E_{0} = A e^{i\phi_{0}}$.
		\While{true}

		\State Forward propagation: $\varTheta_{j} = \mathcal{P}_{\mathrm{MTP}} \left(E_{j} \right) $
		\State Compute OFA intensity: $I ^{\idxn} = \left\lvert \varTheta_{j} ^{\idxn} \right\rvert ^{2}$
		\State Compute intensity accuracy: $ \xi  = 1 - \frac{\left\lVert \widetilde{I}^{\idxn} - I^{\idxn} \right\rVert_{F}}{\left\lVert I^{\idxn} \right\rVert_{F}} $
		\If{$\xi > \mathcal{X} $}
		\Break
		\EndIf
		\State $j = j + 1$
		\State Update weighting factors: $w_{j}^{\idxn} = w_{j-1}^{\idxn} \sqrt{\frac{\widetilde{I}^{\idxn}} {I ^{\idxn}}}$
		\State Update focused field: $\varTheta_{j}^{\idxn} = w_{j}^{\idxn} \sqrt{\widetilde{I}^{\idxn}} e^{i \arg \left[\varTheta_{j-1}^{\idxn}\right]}$
		\State Inverse propagation: $E_{j} = \mathcal{P}_{\mathrm{MTP}} ^ {-1} \left(\varTheta_{j}\right) $
		\State Update CGH: $\phi_{j} = \arg \left(E_{j} \right)$
		\State Update pupil function: $E_{j} = A e^{i \phi_{j}}$

		\EndWhile
	\end{algorithmic}
\end{algorithm}

To quantitatively compare our method with the conventional IFTA, we simulated two OFAs. One has 81 spots with uniform positions but various intensities, and another has 66 spots with random positions and nonuniform intensities. We assumed that the pupil function is modulated by a SLM, and the intensity of the OFA is recorded by a camera. The system parameters of our simulation are listed in Table~\ref{tab:table1}.

\begin{table}[t]
	\vspace{-6pt}
	\centering
	\footnotesize
	\caption{\label{tab:table1}
		\bf System Parameters for Simulation and Experiment.}
	\setlength{\tabcolsep}{0.1\linewidth}
	{
		\begin{tabular}{lc}
			\hline
			\bf{Parameter}                & \bf{Value}         \\
			\hline
			Laser Wavelength (nm)         & 785                \\
			Effective SLM Pixel Number    & $1000 \times 1000$ \\
			SLM Pixel Pitch ($\mu $m)     & 12.5               \\
			Effective Camera Pixel Number & $1001 \times 1001$ \\
			Camera Pixel Pitch ($\mu $m)  & 3.45               \\
			Lens Focal Length (mm)        & 250                \\
			\hline
		\end{tabular}
		\vspace{-6pt}
	}
	\label{tab:performance-comparison-with-equal-weights}
\end{table}

\begin{figure}[!t]
	\centering
	\vspace{-3pt}
	\includegraphics{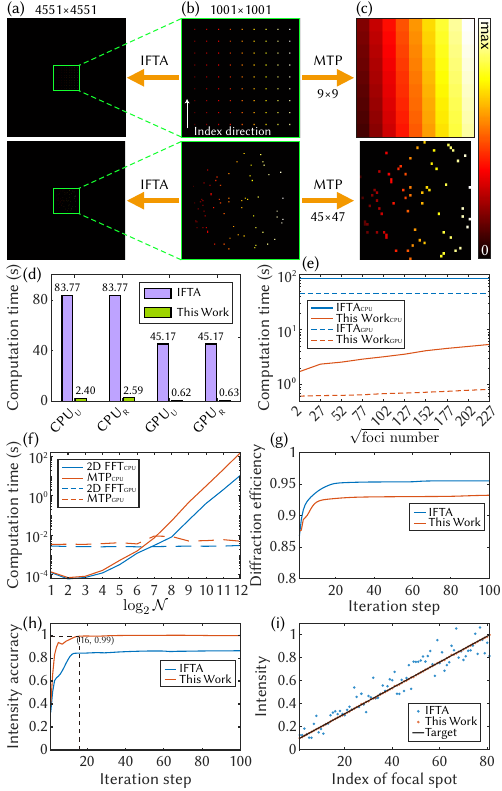}
	\caption{\label{fig:simulated-results} Comparison between conventional IFTA and our approach. \hl{(a)-(c) are the sampling grids of OFAs for the IFTA, camera, and MTP algorithm, respectively. The first and the second rows denote the cases for uniformly and randomly distributed OFAs, respectively. (d) The computation time of the IFTA and MTP algorithm for the two OFAs in 100 iterations. The subscripts U and R denote the results for the uniformly and randomly distributed OFAs, respectively. (e) The computation time of the IFTA and MTP algorithm in 100 iterations, varying with the number of foci. (f) The computation time of the 2D FFT and MTP algorithms in 100 iterations, varying with pixel counts. (g) and (h) are diffraction efficiency and intensity accuracy varying with the iteration step between conventional GS-based IFTA and our method. The diffraction efficiency is defined as the ratio of the energy in the ROI to that of the incident light. (i) is the relative intensity of each focal spot after 100 iterations. Note that (g)-(i) are the results for the generation of the first OFA (uniformly distributed $9 \times 9$ foci).}}
	\vspace{-6pt}
\end{figure}

As mentioned before, the IFTA requires a zero-padding operation of the pupil function. Specifically, this relation follows
\begin{equation}
	M = \frac{M_{0} \lambda}{2 \mathrm{NA} \Delta d},
	\label{eq:FFT-zero-padding}
\end{equation}
where $M_{0}$ and $M$ are the effective sampling number and that after zero-padding along one dimension of the pupil function, respectively. $\Delta d$ is the sampling interval on the focal plane. Therefore, considering the system parameters in Table~\ref{tab:table1}, the sampling number of the pupil function after zero-padding is 4551 for IFTA, which is nearly 5 times more than the effective one.
\hl{Figure~\ref{fig:simulated-results}(b) shows two sampling grids ($1001 \times 1001$) on the camera for uniformly ($1^{\mathrm{st}}$ row) and randomly ($2^{\mathrm{nd}}$ row) distributed OFAs, respectively. To generate such OFAs, the conventional IFTA needs to compute a much larger matrix ($4551 \times 4551$) [Fig.~\ref{fig:simulated-results}(a)], in which the effective sampling only occupies a small region. In contrast, benefiting from the free sampling interval in MTP, the sampling grids on the focal plane can be reduced to $9 \times 9$ [$1^{\mathrm{st}}$ row in Fig.~\ref{fig:simulated-results}(c)] for uniformly distributed OFA and $45 \times 47$ [$2^{\mathrm{nd}}$ row in Fig.~\ref{fig:simulated-results}(c)] for randomly distributed one, respectively.}

The computation time of the two approaches is shown in Fig.~\ref{fig:simulated-results}(d). To reduce the random error, the computation time is counted over 100 iterations and evaluated using the \verb+timeit+ function in MATLAB R2022a. Moreover, the two approaches are evaluated on both CPU (Intel(R) Core(TM) i9-9900KF) and GPU (NVIDIA GeForce GTX 1660). On both platforms, our approach is $>$30-70 times faster than the IFTA. According to Eq.~\ref{eq:FFT-zero-padding}, this acceleration can be further enhanced when a lower NA lens or smaller sampling interval on the focal plane is required. Besides, the spatial distribution of the OFA slightly influences the computation time, indicating that the acceleration of the MTP algorithm mainly comes from the elimination of the zero-padding of the pupil function.
\hl{To validate this, we generate a series of uniformly distributed OFAs with increasing foci number until the scope of the focal plane approaches the upper bound, which is determined by the sampling interval of the pupil plane.
The minimum gap between two foci is assumed to be 20 pixels, therefore the maximum foci number is $227 \times 227$.
The result is illustrated in Fig.~\ref{fig:simulated-results}(e).
It is noted that the computation time of the IFTA keeps constant since the number of sampling points on the focal plane is independent of the foci number.
Although the time cost of our method reasonably increases with the foci number, it is still much faster than the IFTA.
}

\begin{figure*}[!t]
	\centering
	\includegraphics{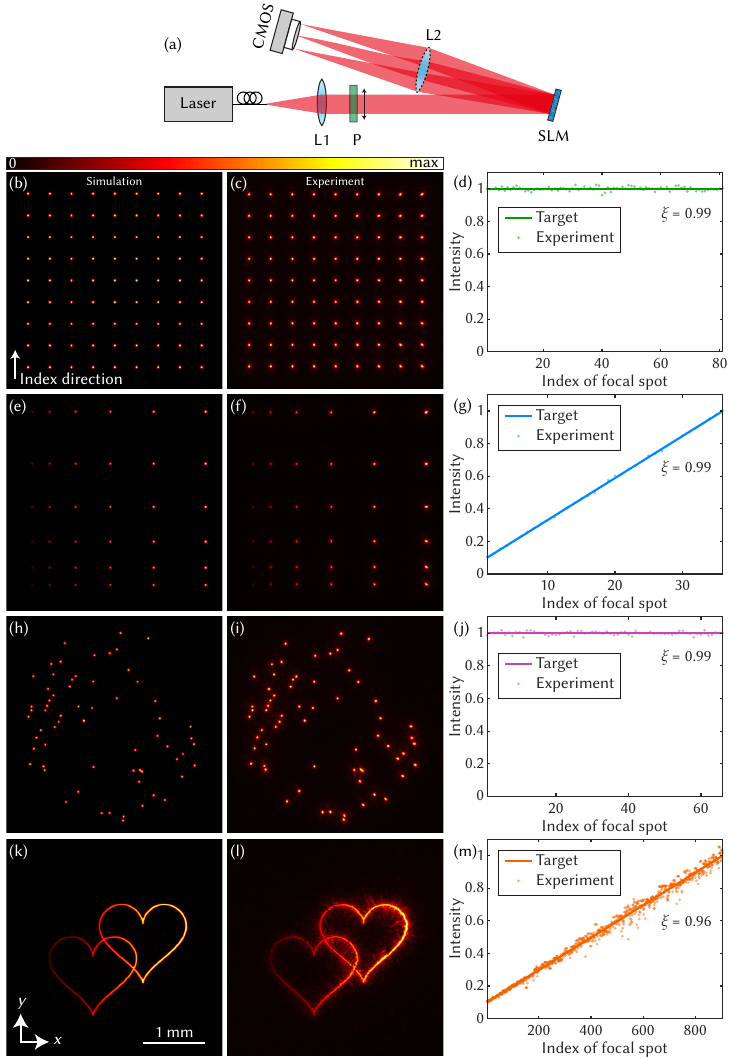}
	\vspace{-3pt}
	\caption{\label{fig:experimental-results} Experimental OFA generation with camera-in-the-loop optimization. (a) Optical setup. Abbreviations: L1 \& L2, lenses; P, polarizer. The double arrow indicates the polarization direction of the laser. \hl{(b) and (c) OFA with uniform positions and intensities of each focal spot in simulation and experiment, respectively. (d) Intensities of each focal spot. (e)-(g) Results for OFA with nonuniform positions and intensities. (h)-(j) Results for OFA with random positions but uniform intensities. (k)-(m) Results for OFA with a double-heart shape and nonuniform continuous intensity distribution.}}
	\vspace{-6pt}
\end{figure*}

Indeed, the computational complexity of the 2D FFT is smaller than that of the MTP when implemented with the same sampling number, $\mathcal{O} \left( \mathcal{N} ^{2} \log_{2}\mathcal{N} \right)$ vs. $\mathcal{O} \left(\mathcal{N} ^{3} \right) $, where $\mathcal{N} $ is the sampling number of the pupil function and the focal plane along one dimension, respectively. However, for the OFA generation, the sampling number of the MTP is much less than that of the 2D FFT. Besides, the matrix product has the potential to be optimized to $\mathcal{O} \left(\mathcal{N} ^{2.4} \right) $~\cite{Zhao:2020:Opt.Lett.} and can be further accelerated by parallel computation techniques.
\hl{If we only assume the same pixel counts of matrices in 2D FFT and MTP, the former is indeed faster [Fig.~\ref{fig:simulated-results}(f)], especially for large pixel counts.
However, as mentioned before, MTP requires no zero-padding and enables flexible samplings, drastically reducing unnecessary computations.}
Therefore, the MTP is more efficient than the conventional 2D FFT for the OFA generation where iterative computation is indispensable.

To compare the quality of the OFA generated by our approach and the conventional IFTA, we adopt the GS-based IFTA as the baseline, since it is also capable of the OFA generation with various intensities of each focal spot. With the generation of the first OFA ($9 \times 9$ spots) as an example, although the diffraction efficiency of the OFA generated by our method is slightly lower than that from the conventional GS-based IFTA, it is still above 90\% [Fig.~\ref{fig:simulated-results}(g)]. The intensity accuracy of the OFA reaches 99\% at the $16^{\mathrm{th}}$ iteration using our weighting strategy, whereas that from the conventional GS-based IFTA even can not reach 90\% after 100 iterations [Fig.~\ref{fig:simulated-results}(h)]. This indicates that, assisted with GPU computation, we can obtain high-quality OFA within 0.1~s. The convergence speed can be further improved using the phase fixing method~\cite{Kim:2019:Opt.Lett.}, but at the cost of lower diffraction efficiency. As shown in Fig.~\ref{fig:simulated-results}(i), the intensities of each focal spot generated using our method correspond well with the target. In contrast, those generated from the conventional GS-based IFTA show large deviations.

Next, we experimentally verified our OFA generation algorithm. The experimental configuration is shown in Fig.~\ref{fig:experimental-results}(a). The laser (Thorlabs, LPS-PM785-FC, 785~nm) is first collimated by a lens (L1: $f = 100$~mm) and then modulated by a polarizer (LBTEK, FLP20-NIR) to ensure the polarization direction is parallel to the orientation of the liquid crystal molecules in the SLM (Hamamatsu, X15223-02, 1024 $\times$ 1272 pixels, 8~bit). This linearly polarized beam impinges on the SLM with a sufficiently small angle. The center of the beam is aligned with that of the SLM. Another lens (L2: $f = 250$~mm) focuses the diffracted beam on a complementary metal-oxide-semiconductor (CMOS) camera (Thorlabs, CS165MU/M, 1080 $\times$ 1440 pixels). The experimental parameters are the same as that of the simulation, listed in Table~\ref{tab:table1}, where the central 1000$\times$1000 and 1001$\times$1001 pixels of the SLM and the camera are used for phase modulation and OFA recording, respectively.

To eliminate the zero-order diffraction and the lens-induced aberrations, we substitute the lens L2 by directly adding an equivalent quadratic phase on the SLM. Moreover, the iterative optimization process cooperates with a camera-in-the-loop strategy, i.e., when the intensity accuracy reaches a threshold (in our implementation, it is 0.95), the calculated OFA intensity is replaced by the captured one for the focused field update. This operation can significantly improve the quality of the OFA since the aberrations and beam profiles are not completely counted in the numerical optimization process.

Firstly, we generated an OFA with uniform positions and intensities of each focal spot [Figs.~\ref{fig:experimental-results}(b),(c)]. The distance between two adjacent focal spots of the experimental OFA is the same as that of the simulated one. Moreover, the intensity accuracy reaches 99\% [Fig.~\ref{fig:experimental-results}(d)]. Figures~\ref{fig:experimental-results}(e)-(g) suggest that our method is also capable of achieving nonuniform OFA. As shown in Figs.~\ref{fig:experimental-results}(h)-(j), for the OFA with random positions of each focal spot, in which two spots may be very close to each other, our method still can handle it well.

Generally speaking, the manipulation of the light at the sub-diffraction-limit scale is more challenging. However, our method is not only valid for OFA generation, which features highly discrete focal spots, but it can be also generalized to continuous pattern generation, i.e, the distance between any two pixels of the target can be within the diffraction limit. As shown in Figs.~\ref{fig:experimental-results}(k),(l), a double-heart with continuous spatial distribution is generated using our method. Remarkably, the intensity accuracy still exceeds 95\% [Fig.~\ref{fig:experimental-results}(m)]. The deviation between these experimental results and the simulated ones may come from many factors, such as the systematic aberrations, the beam profiles, and the modulation precision of the SLM.

\section{Conclusion and outlook}
In summary, we have proposed an approach for the fast generation of arbitrary OFA. Utilizing the MTP for the numerical modeling of the diffraction propagation, the sampling constraints imposed by the FFT in the conventional IFTA have been completely removed, thereby tremendously enhancing the computational efficiency for the generation of arbitrarily arranged focal spots. In addition, the proposed weighting strategy has enabled the generation of OFA with arbitrary intensity assignments of each focal spot after only a few iterations. The simulated and experimental results have shown that both the positions and intensities of each focal spot can be precisely controlled via our approach.

\hl{It should be noted that the proposed method is different from the point-cloud method in computer-generated holography~\cite{Shimobaba:2016:ToII,Tsang:2018:PhotonicsRes.,Blinder:2019:SPIC,Blinder:2022:LightAdvManu}. The point-cloud method describes the OFAs as a group of point light sources. It computes CGH via a directly coherent superposition/integration of waves from every point source. Although it can also be used to compute the wave propagation in OFA generation, the computational complexity is $\mathcal{O} \left(\mathcal{N} ^{4} \right) $, which is much more time-consuming than our method.}

In this paper, some experimental demonstrations are presented. Nevertheless, when questing for a perfect OFA, more practical factors may defect the quality concern. For example, the aberrations should be well corrected, and the beam profile used for iterative optimization should be approximate to that of the practical incident light. Additionally, a well-calibrated SLM is vital for high-quality OFA generation. Although only 2D OFA generation is demonstrated here, our method is also compatible with the 3D case since the axial position of each focal spot can be easily controlled by a defocus phase of the pupil function. As a proof of concept, we generated a 3D visualization of the Chinese character evolution~\cite{liu_xin_2022_6730864}. Moreover, with the development of advanced parallel computation techniques, the computational efficiency of our method can be further improved in the future. Our work offers the potential for those applications where high computational efficiency and flexibility are required.

\section*{Declaration of Competing Interest}
The authors declare that they have no known competing financial interests or personal relationships that could have appeared to influence the work reported in this paper.

\section*{CRediT authorship contribution statement}
\textbf{Xin Liu}: Conceptualization, Methodology, Experiment, Visualization, Writing – original draft. \textbf{Yiwen Hu}: Investigation. \textbf{Shijie Tu}: Investigation. \textbf{Cuifang Kuang}: Supervision. \textbf{Xu Liu}: Supervision. \textbf{Xiang Hao}: Supervision.

\section*{Acknowledgments}
This work was funded by the National Natural Science Foundation of China (92050115), Natural Science Foundation of Zhejiang Province (LZ21F050003), ``Leading Goose'' R\&D Program of Zhejiang (2022C01077), and the Fundamental Research Funds for the Central Universities (226-2022-00137).

% Bibliography
\bibliographystyle{ieeetr}
\bibliography{references}

\end{document}